\documentclass[aip, rsi, twocolumn, showpacs] {revtex4}
\usepackage{graphicx}
\draft 
\graphicspath{{Figs/}}
\begin{document}

\title{Measurement of XeI and XeII velocity in the near exit plane of a low-power Hall effect thruster by light induced fluorescence spectroscopy}

\author{Y.Dancheva, }
\affiliation{CNISM, University of Siena, CSC and DSFTA, via Roma 56, 53100 Siena, Italy}

\author{V.Biancalana}
\affiliation{CNISM, University of Siena, CSC and DIISM, via Roma 56, 53100 Siena, Italy}

\author{D.Pagano, F.Scortecci}
\affiliation{Aerospazio Tecnologie Srl., via Provinciale Nord 42a, 53040 Rapolano Terme (SI), Italy}


\begin{abstract}
Near exit plane non-resonant light induced fluorescence spectroscopy is performed in a Hall effect low-power 
Xenon thruster at discharge voltage of 250~V and anode flow rate of 0.7~mg/sec. Measurement of the axial and radial velocity 
components are performed, exciting the $6s~^2[3/2]^o_2\rightarrow6p~^2[3/2]_2$ transition at  $823.16$~nm in XeI and 
the $5d~[4]_{7/2}\rightarrow6p~[3]^o_{5/2}$ transition at $834.724$~nm in XeII. No significant deviation from the thermal 
velocity is observed for XeI. Two most probable ion velocities are registered at a given position with respect to the thruster
axis, which are mainly attributed  to different areas of creation of ions inside the acceleration channel. 
The spatial resolution of the set-up is limited by the laser beam size (radius of the order of  $0.5$~mm) and the 
fluorescence collection optics, which have a view spot diameter of $8$~mm.
\end{abstract}

\pacs{42.62.Fi; 42.79.Qx; 52.70.-m}

\maketitle 

\section{Introduction}
\label{Int}
Light induced fluorescence (LIF) spectroscopy is a powerful diagnostic technique in thruster characterization.
Velocity measurements based on the Doppler shift of a given species in the plume require accurate and finely 
calibrated laser-based set-ups. Nevertheless, there are unquestionable advantages that make this technique 
very useful. Even though the experimental set-ups are rather complex and delicate, the advanced specifications 
and the robustness of currently available diode laser sources in terms of spectral linewidth, 
frequency and intensity noises, power, and possibility of stable and reliable optical fiber coupling make 
it possible to engineer robust and user-friendly devices. It is worth pointing out the unparalleled properties of  
the LIF velocimetry, which is very unintrusive, has high spatial resolution, offers the possibility to extract information 
even where the plasma itself is created, is species selective, and permits the simultaneous detection of multiple velocity 
components. Excellent results have been obtained with LIF-based techniques in various types of electric thrusters, 
with reference to the 6~kW power range \cite{huang_aiaa_09,huang_jpp_11}, the 1~kW range \cite{gawron_psst_08,bourgeois_pp_10}
and low to medium power range \cite{hargus_jpp_08, hargus_ap_01, cedolin_apb_97}.

The aim of this paper is to characterize a low-power Hall thruster by investigating the behaviour of the 
near exit plane velocity components, using LIF spectroscopy in XeI and XeII. For this purpose a multiple 
channel LIF set-up was built using different free running diode lasers. Details of the laser system, 
the detection/illumination optics, and the spectral calibration system are provided. 
A spectral analysis and the determination of the velocity distribution are presented and discussed.

\section{Apparatus}
\label{App}
	The measurements presented here are performed in a non-magnetic stainless steel vacuum chamber 
  of 1.3~m in diameter and 3~m in length. 	The vacuum pumping is provided by a completely oil free system. 
	In particular, during tests the pumping is performed by a cryogenic system which includes a commercial 
	cryopump and a custom-sized cryopanel cooled to below 50~K by a cryorefrigerator, giving a residual pressure 
	of around $4\times10^{-5}$~mbar (measured with an ionization gauge calibrated for 
	Nitrogen and corrected for Xe using the gas correction factor recommended by the gauge manufacturer).
		
The thruster under investigation is $100\div200$~W class Hall Effect Thruster (HET).  
The external diameter of the acceleration chamber is 40~mm, while the internal diameter is 28~mm. 
The cathode is mounted so that the angle between the thruster output plane and the cathode axis is $45^\circ$. 
The cathode outlet orifice is located approximately 35~mm  radially and 20~mm downstream from the center of 
the channel exit. The thruster is mounted on two electrically driven, micrometric linear stages in order to allow 
it to move in two directions (vertical and horizontal).

\subsection{Diode laser source}
\label{DLS}
\begin{figure}[htbp]
 \includegraphics[width=8cm] {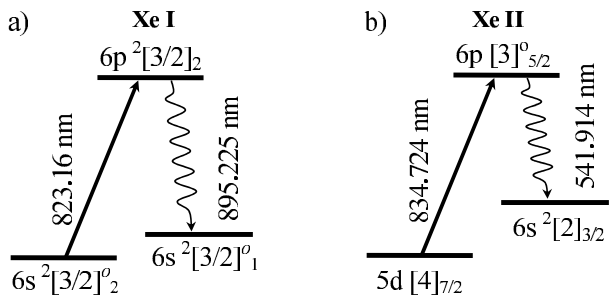}
 \caption{Energy diagrams of the transitions used for non-resonant LIF spectroscopy in a) XeI and b) XeII. The wavelengths are given in air.}
 \label{energy}
 \end{figure}

Non-resonant LIF spectroscopy is performed in XeI and in XeII exciting and detecting the transitions denoted in Fig.\ref{energy}.

\begin{figure}[htbp]
 \includegraphics[width=8cm] {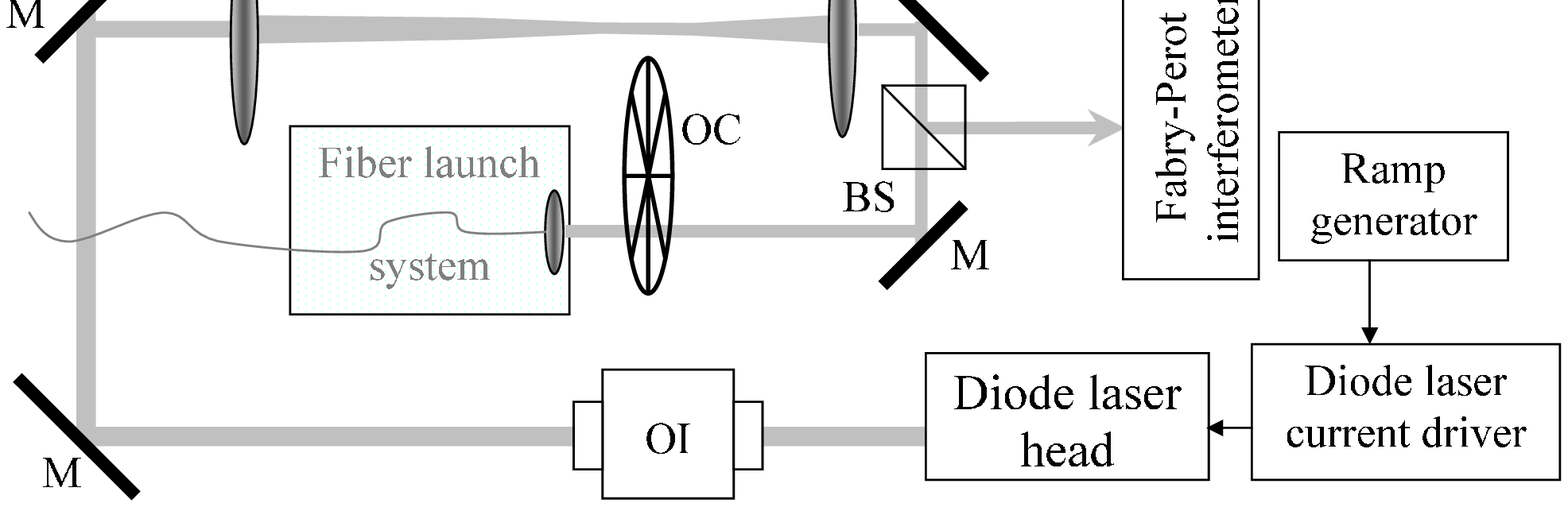}
 \caption{Schematic of the diode laser source: OI-optical isolator; M-mirror; L-lens; OC-optical chopper; BS-beam splitter.}
 \label{diodelasersetup}
 \end{figure}

For this purpose a diode laser system was built using two standard Fabry-Perot index-guided semiconductor chips:
\textit{(i) } laser diode chip (LD1), delivering up to 150~mW of optical power and emitting around 834~nm, and 
\textit{(ii) } LD2, delivering 30~mW and emitting around 823~nm. The diode laser chips are placed in a 
commercial diode laser head and their output beams are launched into a single-mode optical fiber 
(see Fig.\ref{diodelasersetup}). The head allows PID thermostatation with resolution of $0.01^oC$.
The laser beam size is matched to the single mode optical fiber by means of a two-lens optical system, 
followed by a commercial launcher in order to maximize the power coupling. The laser power is 
intensity-modulated in the range of 200~Hz using a mechanical chopper. The fiber coupling efficiency achieved 
is of the order of 50$\%$ and the laser power is subsequently divided into 4 outputs using a 1x4 
single mode fiber splitter. One of the outputs is used to perform spectroscopy in a galvatron spectroscopic lamp
aimed at providing a calibrated-wavelength references. The other three fibers are available for LIF spectroscopy 
of the HET positioned inside the vacuum chamber. Details of the coupling to the vacuum chamber and of the 
illumination system used inside the chamber are reported in Sec.\ref{LCDA}.

Both laser diodes deliver continuous, mode-hop-free tuning in ranges broad enough to perform 
simultaneous spectroscopy in the stationary reference and in the plume of the thruster. Specifically, the LD1
laser can be continuously tuned over more than 40~GHz and the LD2 over more than 15~GHz in the free running regime.

\subsection{Reference signals}
\label{RS}
The atomic/ionic velocities are determined on the basis of the Doppler shift of a 
given known line, measured with respect to a suitable reference line. The choice of the reference 
line must be made considering the frequency separation between it and the expected position 
of the shifted fluorescence line, as this determines the continuous (mode-hop-free) frequency scan 
of the laser diode to be performed. Thus the position of the reference line, together with the 
mode-hop-free detuning of the laser, limits the range of measurable shifts and consequently the
measurable particle velocity. 

In the case of  atomic velocity measurement, the reference line is easily chosen, being 
the same as the unshifted fluorescence line of a stationary discharge, since the velocity of the neutral is 
essentially thermal. In the case of ion velocimetry, and in the absence of an ionic stationary reference line, a 
well-known-wavelength atomic line can be used for reference. In some works (see, for example, \cite{hargus_jpp_08, hargus_jpp_10} 
on LIF velocimetry exciting the ionic transition at 834.724~nm the Doppler shift is determined 
with respect to the atomic line at 834.68217~nm, corresponding to the transition 
$6s'~\left[1/2\right]^o_1\rightarrow6p'~\left\lfloor 3/2\right]_2$ \cite{hargus_jpp_08, lee_phd_10}. Consequently 
diode frequency scans of  the order of 40~GHz must be performed without laser mode-hops. Manzella and colleagues
\cite{manzella_aiaa_94} use the laser back reflection onto the thruster itself as a reference line, which further increases the 
mode-hop-free tuning range required. The unshifted ion line in a stationary discharge is used as well as a reference line, thus
reducing the mode-hop-free range by a factor of 2 (see, for example, \cite{gawron_psst_08}). In this work a different
(more advantageous) atomic reference line is proposed (see details at the end of this section).

\begin{figure}[htbp]
 \includegraphics[width=8.5cm] {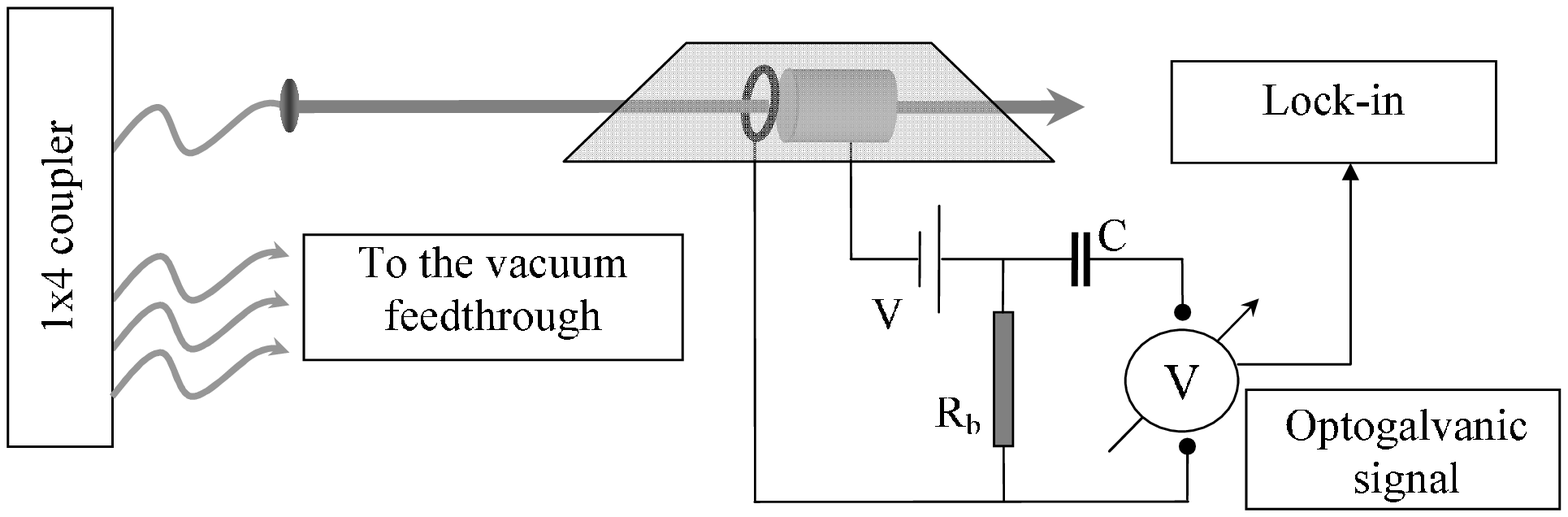}
 \caption{Reference signal arrangement. C=470~nF; Rb=$600~k\Omega$; V=120~V; I=0.1~mA. 
Lock-in is referenced to the optical chopper frequency.}
 \label{referencesignalsetup}
 \end{figure}

The experimental set-up for LIF reference signal detection used in this work is presented in Fig.\ref{referencesignalsetup}. 
As shown, the optogalvanic (OG) signal is recorded as a slight variation in the voltage drop registered at the ballast resistor R$_b$. 
The use of this kind of signal facilitates the task of transduction by making the atomic response signal immediately 
available as an electric one. With a supply voltage of 120~V (around half of it across the discharge lamp) and 
$R_b=600~k\Omega$ the galvatron current is of about 0.1~mA and the optogalvanic signal is of the order of $30~mV_{p-p}$ at 834.68217~nm.

The phase-sensitive detection (PSD) technique is employed for low-noise demodulation of the optogalvanic signal using a 
lock-in amplifier, with a typical integration time of the order of $1~s\div3~s$ (6~dB/oct~$\div$~12~dB/oct output filter). 
As discussed below,  such a long integration time is chosen due to the need to average out the thruster's noise, 
rather than the optogalvanic lamp discharge noise.

Continuous frequency tuning is performed using a triangular wave generator applied to the modulation input of the 
diode laser current driver. A maximum $1~V_{p-p}$ triangular wave signal is applied to the diode laser current driver input, 
causing a driving current variation of 20~mA. Its frequency is very low (of  the order of few mHz) so that the junction 
current is varied slowly  with respect to the measuring time. According to the laser-driver modulation conversion 
factor (input modulation voltage to junction current) and to the laser diode chip response (junction current to optical 
frequency), the scan produces a continuous sweep of the optical frequency. This sweep 
is accurately monitored with the help of a confocal Fabry-Perot (FP) interferometer with a free spectral range of 1.47~GHz. 
The signals from the FP cavity and from the reference lamp are  acquired simultaneously, with the LIF signal 
being demodulated by a second lock-in amplifier, so that they provide a precise relative (the FP) and absolute (the stationary discharge) 
calibration of the frequency scale. 

\begin{figure}[htbp]
 \includegraphics[width=8cm] {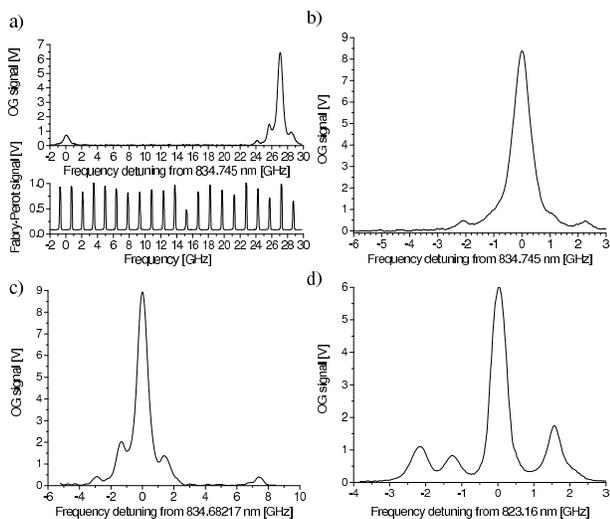}
 \caption{Reference optogalvanic signals at: b) 834.745~nm c) 834.68217~nm d) 823.16 nm and a) at both  834.68217~nm 
and 834.745~nm together with the Fabry-Perot frequency comb.}
 \label{referencesignals}
 \end{figure}

Examples of the reference curves are shown in Fig.\ref{referencesignals}. The laser power illuminating the discharge 
lamp is attenuated down to 2~mW using a neutral density filter. 
Fig.\ref{referencesignals}  shows the relevant lines available for the frequency scale-offset calibration and includes (plot a) the
Fabry-Perot comb of peaks used for frequency scale-factor calibration. Besides the reference line, at 823.16~nm for XeI velocimetry (plot d) and 
at 834.68217~nm for XeII velocimetry (plot c), the XeI has a weaker but clearly detectable line at 834.745~nm 
(transition $6s~^2[3/2]_1\rightarrow8s~^2[3/2]^o_1$ ). This last line (plot b) is used as a reference line for XeII velocimetry in this work,
as it is 9~GHz red-shifted (the Doppler shift of the axial velocity of the moving ionic particles is also red) with respect to the 
stationary ionic one (at 834.724~nm). The fact that this reference line is closer to the shifted ionic one decreases  the 
necessary frequency tuning range by an other factor of 2, thus speeding up the measurement time. 
The weaker amplitude is not problematic thanks to the good signal-to-noise ratio given by the OG signal.

\subsection{Light collection arrangement and data acquisition}
\label{LCDA}
Two of the outputs of the $1\times4$ fiber splitter are coupled to illuminators placed inside the vacuum chamber using 
a four-port vacuum feedthrough. The optical power losses through the vacuum feedthroughs are significant for three out of four.
Power loss levels in excess of 70~\% are measured for three feedthroughs and 30~\% for the other one. Inside the vacuum chamber, 
high-vacuum  single mode fibers are used. The laser beam, applied in the axial ($\hat z$) direction, is collimated using an adjustable 
focal length collimator, providing a beam waist of 0.8~mm (diameter at $1/e^2$ intensity) at a distance of 1 m 
(the distance between the fiber output and the thruster output plane). The fluorescence signal is collected using a multimode 
fiber equipped with collimation optics (22~mm clear aperture (CA) collimator), which determines the lowest spatial selectivity 
(the highest view spot diameter). The view spot diameter is adjusted using a remotely controlled multiple-diameter diaphragm 
ranging from 18~mm to 8~mm, placed in the proximity of the collimator entrance.

The collected light  is extracted from the vacuum chamber through a multi mode fiber vacuum feedthrough and is directed 
to a 0.5~m monochromator. The monochromator slits are adjusted to set maximum transmittance at 895.22~nm for LIF in XeI, 
and at  541.914~nm for LIF in XeII, with a spectral  width of about 1.5~nm, in order to perform non-resonant LIF 
spectroscopy and improve the signal-to-noise ratio by rejecting broad spectrum radiation. This spectral width abundantly
exceeds that of the diode laser and the LIF signal, so that the monochromator acts as an adjustable band-pass filter
with a flat-response in the range of interest and has no effect on the spectral features measured. A high gain, 
ultra-low noise photomultiplier tube (PMT) is used as a detector at the monochromator output. 

Four signals are acquired simultaneously: the triangular wave signal applied to the diode laser current driver; the 
Fabry-Perot fringes; the optogalvanic signal, and the LIF signal, both at the output of the lock-in amplifiers. The ramp 
signal is used to determine the direction of the frequency ramp. The Fabry-Perot signal is used to linearize the 
diode laser frequency scan and to establish a calibrated relative frequency scale. The optogalvanic signal is used to determine 
the frequency shift with respect to the atomic line nearest to the ionic one and thus to calibrate the frequency scale offset. 
Multiple measurements are performed at each $\hat y$, $\hat z$ position. This helps to improve the
signal-to-noise ratio by averaging and provides uncertainty estimation in terms of the sample standard deviation of the
inferred velocity.

\section{Results}
\label{Results}
\subsection{LIF velocimetry in XeI}
\label{LIFXe_atom}
The performance, lifetime and operational stability of a HET depend on its neutral propellant flow parameters.
For example, the azimuthal uniformity of the number density at the propellant distributor exit plane
has been studied in detail \cite{baranov_iepc_01, reid_phd_09, langendorf_rsi_13} as an
important parameter to obtain efficient performance of the thruster.

\begin{figure}[htbp]
 \includegraphics[width=7.5cm] {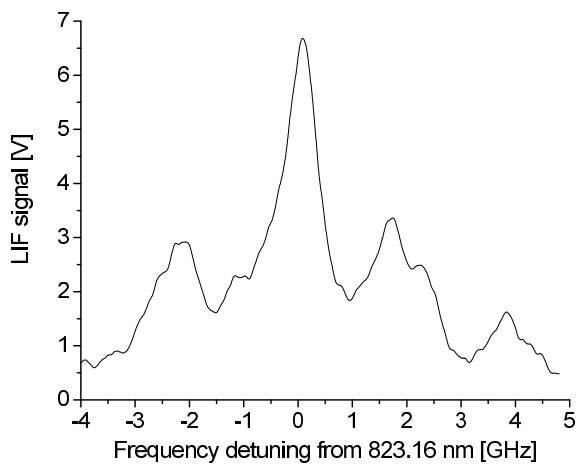}
 \caption{LIF signal at 895~nm for the HET at: V=250~V; I=0.8~A; Xe anode flow 0.8~mg/sec;
lock-in $\tau=3$~sec; (output filter at 12~dB/oct); laser power of 1.6~mW in axial direction.}
 \label{fig:LIFXeI}
 \end{figure}

In this work the axial velocity component of Xe atoms is only recorded for the purpose of
illustrating the set-up's potential for multiple species interrogation. The thruster channel is illuminated 
at 823~nm (laser power of 1.6~mW) and the fluorescence is collected at a distance of $z$=-20~mm 
from the thruster output plane, with a collimator view point diameter of 18~mm. The LIF signal at 895~nm 
is presented in Fig.\ref{fig:LIFXeI} (for a Xe anode flow of 0.8~mg/sec). The spectrum of the signal is determined
by the multiple Xe isotopes and the hyperfine structure (see for comparison Fig.\ref{referencesignals}d). The single axial 
velocity component was found to be essentially thermal. Similar results have been obtained for Xe anode flows down to 0.6~mg/sec.
The  small negative axial component measured, of about 30~m/sec, was attributed to the anode temperature 
increase and/or collisions with Xe ions.

\subsection{LIF velocimetry in XeII}
\label{LIFXeII}
\begin{figure}[htbp]
 \includegraphics[width=8cm] {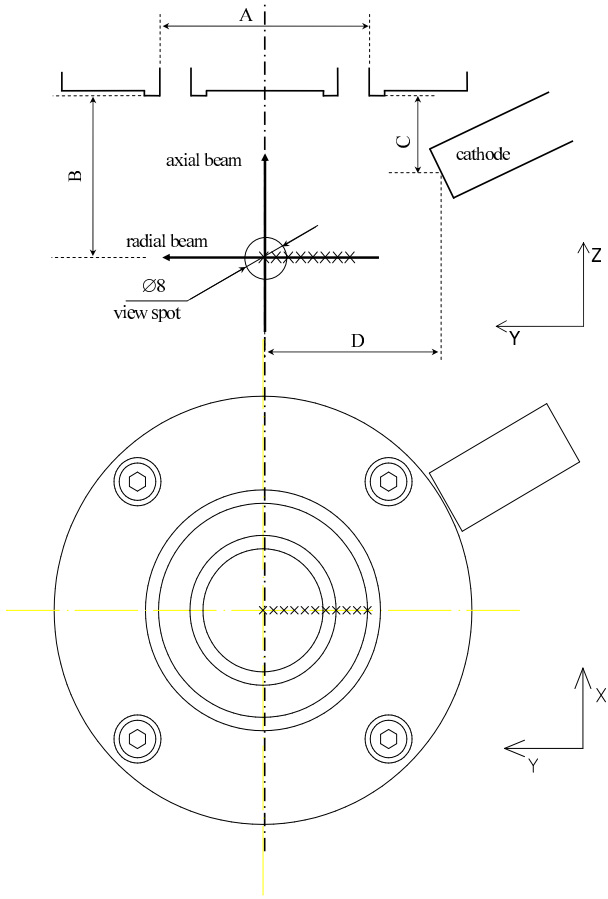}
  \caption{Top and front view of the experimental arrangement inside the vacuum chamber 
	(A=40~mm, B=37~mm, C=17~mm, D=35~mm). The 
	position of the view spot is presented with respect to the thruster output plane. The various y
	positions of the center of the thruster with respect to the illumination system are 	marked with $\times$. 
	The origin of the co-ordinate system is positioned at the center the thruster.}
 \label{OS_vacuumchamber}
 \end{figure}

Two velocity component measurements were carried out in order to determine ion velocity along the two orthogonal 
directions (axial and transverse directions). One laser beam, with a power of 8.5~mW and diameter of 0.8~mm (at the thruster plane), 
is used to determine the velocity component along the direction perpendicular to the thruster exit plane. 
A second laser beam, slightly larger in diameter (1~mm) and lower in power (3~mW), travels along the transverse 
direction, providing the radial component of the velocity if the beam lies on one of the symmetry planes 
of the thruster (approximated with a cylinder if the presence of the cathode is neglected geometrically).
The schematic of the laser illumination system set-up inside the vacuum chamber is sketched in 
Fig.\ref{OS_vacuumchamber}, together with the location of the view spot with respect to the thruster 
output plane and the cathode position. The spontaneous and laser-induced fluorescence is collected with the multimode 
optical fiber (positioned at around $x$=500~mm from the thruster) through the 8~mm diaphragm (see Fig.\ref{OS_vacuumchamber}). 
The intersection of the laser interaction volume with the view spot forms the light-plume detection volume, of less than 1x1x8~mm.

Table~1 gives the nominal thruster operating conditions. 
\begin{table}
\begin{tabular}{||l|l||}
\hline
 Anode mass flow rate & 0.7~mg/sec \\
\hline
Cathode mass flow rate & 0.2~mg/sec \hfill \\
\hline
 Anode potential  & 250~V \hfill \\
\hline
Anode current  &  0.67~A  \hfill \\
\hline
\end{tabular}
\caption{Nominal thruster operating conditions}
\label{T1}
\end{table}

A two-axis translation system makes it possible to move the thruster in the ($\hat x$,$\hat y$) plane with respect to the 
fixed measurement optical system. The axial and transverse velocity components can be measured on all the 
points lying on this plane parallel to the thruster exit plane and 37~mm from it, i.e. 20~mm beyond the cathode position.

\begin{figure}[htbp]
 \includegraphics[width=8cm] {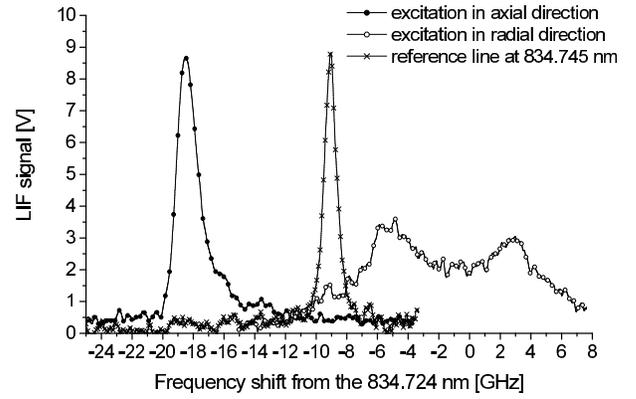}
 \caption{Examples of the LIF signals (at y=0) for laser excitation in both axial and radial directions, in scale.}
 \label{LIFexamples}
 \end{figure}

Both laser beams (applied in axial and radial directions) are aligned so as to cross at the center of the view spot. 
Due to the noise coming from the thruster itself, a long time constant must be used for LIF phase sensitive detection
and thus no strict requirements are imposed for the diode laser intensity modulation frequency (using the mechanical 
chopper). As mentioned in Sec.\ref{RS}, the diode laser optical frequency is scanned slowly using a triangular wave generator. LIF 
signals such as those presented in Fig.\ref{LIFexamples} can be registered using a lock-in amplifier (we use a Stanford SR830 
lock-in amplifier) with a time constant of 1~sec and a 12~dB/oct output filter. The integration time set in conjunction 
with the diode laser frequency sweep width makes it necessary to use triangular wave signals with a frequency of the order of 
2~mHz and thus a measurement time of the order of 4~min (one scan).

\begin{figure}[htbp]
 \includegraphics[width=8cm] {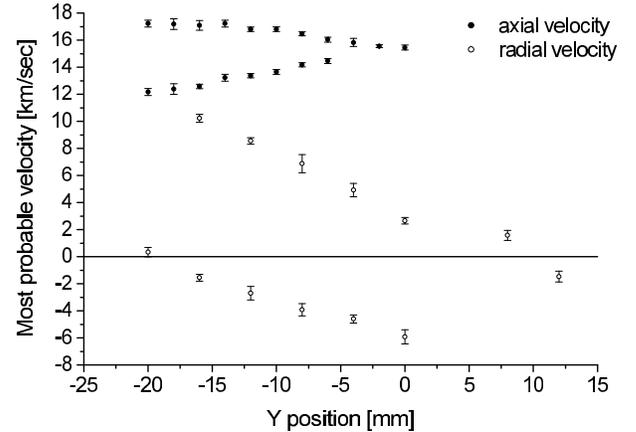}
 \caption{Axial and radial components of the XeII velocity as a function of the $\hat y$ position. The error bars are the 
mean-square-root deviation over several measurements.}
 \label{mappa}
 \end{figure}

With our experimental parameters the contrast of the LIF signal (determined as the ratio between the ac rms amplitude given by the lock-in 
and the PMT dc signal) is of the order of 1~\% . The axial LIF signal presents a single, red-shifted peak at 18.5~GHz, 
corresponding to the most probable velocity of 15.4~km/sec (calculated as $\vartheta=\Delta\nu/\lambda$ and corresponding 
to $\approx163$~eV) in the opposite to the laser beam direction and for y=0. Two well-separated components of the radial velocity 
are observed at the center of the thruster, as well as at any displacement in the $\hat{y}$  direction (see Fig.\ref{mappa}). 
The radial non-zero velocity components show quite asymmetric peaks at 24~eV (radial component in the direction of 
the cathode) and nearly 5~eV (in the opposite direction with respect to the cathode). 

The fluorescence spectrum changes in terms of both shape and peaks positions as it moves away from the center of the 
thruster exit plane, reflecting the decrease of the ion density and the variation in the distribution of the ion velocity. 
The presence of double axial and radial velocity components is registered for y$\geq$-8~mm.
Double radial velocity components at y=0 are likely due to overlapping of the ion flows coming 
from opposite sides of the accelerating channel. The presence of multiple velocity components is also 
found  for the axial component. Fig.\ref{mappa} synthetically represents this behaviour, 
showing the most probable axial and radial velocity as a function of the $\hat y$ position. In particular,  
concerning the axial velocity, a single peak is present up to y=-4~mm from the center along the $\hat y$  direction  
the peak width increases progressively. At y=-6~mm a second peak appears, not yet well separated from the first one.
Beyond this distance, a second peak becomes clearly visible.

Concerning the axial velocity, a single velocity peak is observed for only 5~mm away from the center of the thruster. 
The broadening of the peak towards the lower velocities (corresponding to a decrease in the mean velocity) and subsequent appearance of 
two most probable velocities occur when the thruster in displaced in the $\hat y$  direction. 
It should be pointed out that the relative intensity of the two axial peaks is comparable.

The distribution behaviour of the radial velocity as a function of the radial displacement is different. There is a big discrepancy between the 
most probable velocity and the mean velocity. The mean radial velocity value as a function of the $\hat y$  displacement is close 
to zero. The radial LIF signal appears as a broad \textit{plateau} with two side peaks of different intensities at distances bigger than $y$=-4~mm. 
When both are present,  the intensity ratio between the two most probable radial velocities is of the order of 2. 

\section{Conclusions}
\label{Conclusions}
	A  set-up was built to measure multiple velocity components in XeI and XeII, based on Doppler shift measurements
	of LIF signals. The good spatial resolution permitted detailed investigation of the plume at centimetric distances 
	from the exit plane of the thruster. Both XeI and XeII velocities were measured. No significant discrepancy 
	from a thermal velocity was found in the case of XeI. In the case of XeII, and at a distance of $z$=-37~mm, a map 
	of the axial and radial velocity could be drawn and it was determined that XeII is accelerated to approximately 
	163~eV at the center of the thruster. The poorer radial spatial selectivity with respect to the axial one made it 
	difficult to establish whether the 24~eV peak contributed to the maximum energy together with the axial peak.
	
	A multiple channel LIF spectroscopy set-up was realized with the help of optical fibers. The system offers LIF spectroscopy
	(and thus accurate velocimetry) in both XeI and XeII by changing only the diode laser head. The possibility of 
	using the transition at 834.745~nm as a reference was demonstrated, bringing the advantage of narrower
	diode laser frequency scanning, thus	shortening the measuring time with respect to other reference lines 
	and relaxing the requirements in terms of laser mode-hop-free	tuning.

\section{Acknowledgments}
\label{Acknowledgments}
The research  was co-financed by Tuscany's Regional administration under the regional R\&D program, 
POR CreO FESR 2007-2013 Bando Unico R\&S 2008, whose support is gratefully acknowledged. 
The authors would like to acknowledge Dr. Giovanni Coduti for his help in building the experimental 
set-up, Mr. Simone Scaranzin for the tests carried out at Aerospazio Tecnologie Srl, and thank Emma Thorley 
for reviewing the manuscript. Y.D. and V.B. thank C.Stanghini and L.Stiaccini for their valuable technical support.


\end{document}